\documentclass{emulateapj}
\usepackage{apjfonts}
\usepackage{epsf}
\usepackage{graphicx}


\lefthead{INOUE and SILK}
\righthead{Local Voids as the Origin of Large-angle CMB Anomalies I}

\newcommand{\K}{{\mathbf{k}}}
\newcommand{\x}{{\mathbf{x}}}

\newcommand{\f}{\frac}

\newcommand{\bb}{\bibitem}
\newcommand{\BF}{\begin{figure}\begin{center}}
\newcommand{\EF}{\end{center}\end{figure}}
\newcommand{\BE}{\begin{equation}}
\newcommand{\EE}{\end{equation}}
\newcommand{\BEA}{\begin{eqnarray}}
\newcommand{\EEA}{\end{eqnarray}}
\newcommand{\ti}{\textit}

\begin{document}
\title{Local Voids as the Origin of Large-angle Cosmic Microwave Background Anomalies I}
\author{Kaiki Taro Inoue  \altaffilmark{1} and Joseph Silk
\altaffilmark{2}}

\altaffiltext{1}{Department of Science and Engineering, 
Kinki University, Higashi-Osaka, 577-8502, Japan}
\altaffiltext{2}{University of Oxford, Department of Physics, 
Oxford, OX1 3RH, United Kingdom }

\begin{abstract}
We explore the large angular scale temperature anisotropies 
in the cosmic microwave background 
due  to expanding homogeneous local voids at redshift $z\lesssim 1$.
A  compensated  spherically symmetric homogeneous dust-filled 
void  with  radius 
$\sim 3\times 10^2 h^{-1}$Mpc, 
and  density contrast $\delta \sim -0.3$ can be 
observed as a cold spot with a temperature anisotropy
$\Delta T/T\sim -1\times 10^{-5}$ surrounded by a 
slightly hotter ring. We find that a pair of these circular 
cold spots separated by $\sim
 50^\circ$ can account both for the 
planarity of the octopole and
the alignment between the quadrupole and the octopole in the
cosmic microwave background (CMB) anisotropy. The 
cold spot in the Galactic southern hemisphere which is anomalous at the 
$\sim 3\sigma$ level can  be explained by such a large 
void at $z\sim 1$.    
 The observed north-south asymmetry in
the large-angle CMB power can be attributed to the 
asymmetric distribution of these local voids between the 
two hemispheres. The statistical significance of the 
low quadrupole is  further reduced in this interpretation of the 
large angular scale CMB anomalies.
\end{abstract}
\keywords{cosmic microwave background -- cosmology -- large scale structure}

\section{introduction}
Despite the success of the cold dark matter ($\Lambda$CDM) model
in explaining the 
cosmic microwave background (CMB) anisotropy 
on  small angular scales, the possible presence of 
anomalies on large angular scales in the 
Wilkinson Microwave Anisotropy Probe 
(WMAP; Bennett et al. 2003) data  
has triggered a debate over their origin. 

The reported anomalies are as follows: 
the anomalously low quadrupole observed by COBE/DMR (Smoot et al. 1992)
and WMAP (Bennett et al. 2003);
the octopole planarity and the alignment between the quadrupole
and the octopole (Tegmark et al. 2003; de Oliveira-Costa et al. 2004); 
an anomalously cold spot on angular scales 
$\sim 10^\circ$ (Vielva et al. 2004;Cruz et al. 2005, 2006);
and an asymmetry in the large-angle power
between  opposite hemispheres (Eriksen et al. 2004; 
Hansen et al. 2004). Evidence for other forms of 
non-Gaussianity on  large angular scales has also 
been reported (Chiang et al. 2004; Park 2004;
Schwarz et al. 2004, Larson \& Wandelt 2004). It should of course be emphasized that 
the significance of any one of these anomalies
is at most $3~\sigma.$ Nevertheless, the accumulation of anomalies
suggests that there may be hints of new physics to be added to the standard 
cosmological model. The possible implications are sufficiently 
important that discussion of theoretical explanations is justified. 

Indeed, to explain the origin of the anomalies,  various 
solutions have been suggested. Luminet et al.(2003) proposed 
a non-trivial spherical topology to explain the 
low quadrupole, and Jaffe et al.(2005) considered a locally anisotropic 
model based on the Bianchi type $\textrm{VII}_h$ universe 
to explain the quadrupole/octopole planarity and alignment. 
Other papers have studied  the possibilities that the large-angle CMB
is affected by local non-linear inhomogeneities (Moffat 2005; 
Tomita 2005a, 2005b; Vale 2005; Cooray \& Seto 2005; Raki\'c et al. 2006).
None of these explanations is  fully satisfactory, and 
it is therefore useful to develop an alternative  interpretation of the  anomalies.

In this paper, we explore the possibility that the CMB is affected by  
a small number of local voids at redshifts $z\lesssim 1$.
As is well known, the Rees-Sciama effect 
(Rees \& Sciama 1968) for voids 
could induce an additional anisotropy on large angular scales.
Thompson \& Vishniac (1987) and Mart\'inez-Gonz\'alez \& Sanz (1990)
have studied the anisotropy owing to a vacuum 
spherical (Swiss-cheese) void using the thin-shell approximation.
Panek (1992), Arnau et al. (1993), and Fullana et al. (1996)
studied voids for a general spherically symmetric energy profile
based on the Tolman-Bondi solution (TBS;Tolman 1934; Bondi 1947). 
Although the TBS is a good description of 
voids with a general spherically symmetric density profile,
it has a drawback in that it only applies before shell crossing.
In contrast, the thin-shell approximation is suitable
for studying voids in the late stages of evolution. In section 2, 
we first derive  analytic formulae for  dust-filled voids
in the Einstein-de Sitter background using the thin-shell approximation.
In section 3, we explore a  model of local voids that agrees 
with the observed anomalies on large angular scales.  
In section 4, we summarize our results.
\section{Dust-filled void model}
\vspace{0.2cm}

\subsection{Cosmic expansion}
We estimate the temperature anisotropy caused 
by a homogeneous thin-shell void.  
For simplicity, we consider 
a matter-dominated Friedmann-Robertson-Walker
(FRW) model with zero curvature (Einstein-de-Sitter model) 
as the background universe. 
In what follows, we use 
units in which the light velocity $c$ is normalized to 1. 
In
spherical coordinates $(r,\theta,\phi)$, the flat background FRW metric
can be written as 
\BE
ds^2=-dt^2+R^2(t)(dr^2+r^2 d \Omega^2),
\EE
where $t$ is the time and $R$ denotes the scale factor, and
$d \Omega^2$ is the metric for a unit sphere.
Next, we consider a 
homogeneous spherical dust-filled void 
with a density contrast $\delta<0$. 
We assume that the size of the void is 
sufficiently smaller than the Hubble radius $H^{-1}$. Then, the metric
of the void 
centered on the origin can be approximately described by 
the hyperbolic FRW metric as 
\BE
ds^2=-dt'^2+R'^2(t')(dr'^2+ R_c^2 \sinh^2(r'/R_c)d \Omega^2),
\EE
where $t'$ is the time, $R'$ denotes the scale factor, 
and $R_c$ is the comoving
curvature radius. Assuming that the curvature radius $R_c$
is sufficiently smaller than the void radius $r_v$, up to 
order $O((H'^{-1}/R_c)^2)$, the scale factor 
for the hyperbolic FRW void can be written as 
\BE
R'(t') \approx \biggl(\f{3}{2}\biggr)^{2/3} 
(\Omega' H'^2)^{1/3}t'^{2/3} 
+ 3\biggl(\f{3}{2}\biggr)^{1/3}\f{(\Omega' H'^2)^{-1/3}}{10
R_c^2}t'^{4/3}. \label{eq:fac1}
\EE 
In terms of the Hubble parameter contrast 
$\delta_H\equiv H'/H-1$ and the density contrast
$\delta=\rho'/\rho-1$, equation (\ref{eq:fac1}) is
expressed as  
\BE
R'(t')\approx (1+\delta)^{1/3}\biggl(\f{t'}{t}\biggr)^{2/3} + 
 \f{\kappa}
{5(1+\delta)^{1/3}}\biggl(\f{t'}{t}\biggr)^{4/3}, 
\label{scale}
\EE 
where $\kappa \equiv H^{-2}/(R'^2 R_c^2)=(1+\delta_H)^2-(1+\delta)$
is the absolute value of the 
curvature in unit of the background 
Hubble parameter $H$. 
Up to order $O(\kappa)$, equation (\ref{scale}) 
yields the cosmic age for the void
in terms of the model parameters $\delta_H$ and $\kappa$ 
given at a specific epoch as
\BE
t'=t \biggl (\f{1}{1+\delta_H}+ \f{\kappa}{5(1+\delta_H)^3}
\biggr). \label{t'}
\EE

\subsection{ Time solution}

In what follows, we calculate the evolution of the internal 
time $t'$ in terms of the external time $t$ for the expanding 
void (with a peculiar velocity) up to 
order $O((r_v/H^{-1})^3)$. The comoving radius of the 
void is expressed as $r_v(t)$ in the external coordinates
and as $r'_v(t')$ in the internal coordinates.
To connect the two metrics at the shell,  we require
the following boundary conditions:
\BE
(R(t)r_v(t))^2=(R'(t')r'_v(t'))^2\biggl(1+\f{R'(t')^2 r'_v(t')^2}{3 R_c^2}\biggr), 
\label{con1}
\EE
and
\BE
-dt^2+R^2(t)dr^2=-dt'^2+R'^2(t')dr'^2. \label{con2}
\EE
where we have assumed that $r'\ll R_c$.
Up to order $O((r_v/H^{-1})^3)$, the curvature term in 
equation (\ref{con1}) is negligible. Then equation 
(\ref{con1}) and  (\ref{con2}) yield
\BE
dt'=dt\biggl
(1-R\dot{R}r_v\dot{r_v} \delta_H+\f{1}{2}\dot{R}^2 r_v^2 \delta_H^2
\biggr), \label{dt'1}
\EE
where a dot means the time derivative $d/dt$.
Let us assume that the expansion of the thin shell
in the external coordinates is expressed as $r_v(t)\propto t^\beta$, where $\beta$
is a constant. For the Einstein-de Sitter background, the matter
dominated universe expands as $R(t)\propto t^{2/3}$. Therefore, 
equation (\ref{dt'1}) can be explicitly written as 
\BE
dt'=dt\biggl
(1-\f{2}{3} \beta \delta_H \eta^2+\f{2}{9} \delta_H^2 \eta^2 
\biggr),\label{dt'2}
\EE
where we define 
\BE
\eta \equiv \f{r_v(t) R(t)}{t}. \label{eta}
\EE
Integrating equation (\ref{dt'2}), we find
\BE
t'\approx   
t\biggl(1-\f{2 \beta}{6 \beta+1}\delta_H \eta^2
+\f{2}{18 \beta+3}\delta_H^2 \eta^2\biggr).
\EE \label{Deltat'}
Note that $\delta_H$ and $\eta$ are functions 
of $t$ rather than $t'$. As we see in the 
subsequent analysis, the anisotropy turns out 
to have a leading order of $\eta^3$. Therefore, 
omitting the curvature term in equation ($\ref{con1}$)
\textit{in the process of deriving} $dt'$ can be justified.

\subsection{Expansion of voids}
In the Newtonian limit, the peculiar velocity of the void 
$v$ can be 
written in terms of the gravitational acceleration $g$ and
the Hubble parameter $H$ and the density parameter 
$\Omega$ as
\BE
v=\f{2 f g}{3 H \Omega}, \label{v1}
\EE
where $f$ is written in terms of the 
linear growth factor $D$ and the scale factor $R$ as
\BE
f=\f{R}{D} \f{d D}{d R}.\label{f1}
\EE
Equation (\ref{f1}) can be 
approximately written as (Peebles 1980)
\BE
f(\Omega)\sim \Omega^{0.6}.\label{f2}
\EE
Consider a homogeneous void with an inner radius $r_-$ and an outer 
radius $r_+$. We assume that the mass deficiency
inside the void $-m $ is equal to the mass of the wall
if the contribution from the mean density is subtracted.
If we assume further 
that the motion of the wall is described by the 
distance $d_*$ from the center of the void 
where the gravitational mass satisfies $M(d_*)=m/2$,
the gravitational acceleration $g$ acting on the wall is
\BE
g=-\f{G (m/2)}{d_*^2},~~ m=\f{4 \pi }{3}d_*^3 \delta \rho,
\label{g}
\EE
where $\rho$ denotes the background density. 
Plugging $g$ in equation (\ref{g}) 
into equation (\ref{v1}), (\ref{f1}), and  (\ref{f2}),
we have (Sakai, 1995, Sakai et al. 1999),
\BE
v =-\f{1}{6}H d_*\Omega^{0.6} \delta. \label{velocity}
\EE
Assuming $d_*\sim r_v$, the power of the expansion for 
a compensating void is 
\BE
\beta\approx -\f{1}{6}\Omega^{0.6} \delta. \label{beta}
\EE

\subsection{Temperature anisotropy}
We denote quantities at the time  the photon enters
the void and those at the time  the photon leaves
by the subscripts ``1'' and ``2'', respectively.
Primes denote quantities measured by a comoving observer
in the interior coordinate system and the unprimed quantities
are measured by a comoving observer in the background universe
just out of the shell of the void. The angles $\psi_1$, 
$\psi'_1$, $\psi'_2$, and $\psi_2$ are defined by reference to 
figure 1. 

\begin{figure*}
\epsscale{1}
\plotone{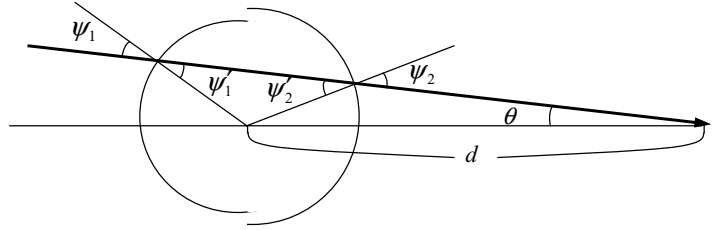}
\vspace{-3cm}
\caption{Cross section of a spherical void through the plane in which 
the photon path and the line of sight to the void center lie. 
Here $\Psi$ is the angle between the light ray and
the normal vector on the spherical void.
The subscripts 1 and 2 represent the quantities at the time
the photon enters and leaves, respectively. Primed variables 
represent the values inside the void.}
\label{cross-section-void.eps}
\end{figure*}
To calculate the energy loss, we apply two local
Lorentz transformations at each void boundary. 
The first is to convert the photon four-vector momentum 
in the comoving frame in the background universe
to the frame in which the shell is at rest. 
The second is to convert it to the frame in the comoving frame
inside the void.

The  four-vector momentum of the photon that enters the void is
\BE
\bold{k}_1 \equiv E_1\left(
\begin{array}{c}
1 \\ \cos\psi_1 \\ \sin\psi_1 \\ 0
\end{array}\right),
\EE
where $\psi_1$ is the angle between the normal vector of the void shell 
and the spatial three-vector of the momentum. 
After the photon passed the shell,
the four-vector is converted to 
\BEA
\bold{k}'_1 
&\equiv&
 E_1'\left(
\begin{array}{c}
1 \\ \cos\psi'_1 \\ \sin\psi'_1 \\ 0
\end{array}
\right)
\\
&=& E_1\left(
\begin{array}{c}
\gamma_1\gamma_1'[1+(v_1-v_1')\cos\psi_1-v_1v_1'] \\
\gamma_1\gamma_1'[\cos\psi_1+(v_1-v_1')-v_1v_1'\cos\psi_1] \\
\sin\psi_1 \\ 0
\end{array}\right), \label{k1'}
\EEA
where $v_1$ and $v_1'$ are the velocities of the void shell
at the time $t=t_1$ and $\gamma$ factors are defined as 
$\gamma_1=1/(1-v_1^2)^{1/2}$ and $\gamma_1'=1/(1-v_1'^2)^{1/2}$.  
When the photon reaches the far edge of the shell, the 
four-vector momentum becomes 
\BE
\bold{k}_2' \equiv \f{R_2'}{R_1'}E_1' \left(
\begin{array}{c}
1 \\ \cos\psi_2' \\ \sin\psi_2' \\ 0
\end{array}\right).
\EE
As the photon leaves the shell, the four-vector momentum
is converted to 
\BEA
\bold{k}_2 
&\equiv&
 E_2\left(
\begin{array}{c}
1 \\ \cos\psi_2 \\ \sin\psi_2 \\ 0
\end{array}
\right)
\\
&=&
\f{R_2'}{R_1'}E_1'\left(
\begin{array}{c}
\gamma_2\gamma_2'[1+(v_2-v_2')\cos\psi_2'-v_2v_2'] \\
\gamma_2\gamma_2'[\cos\psi_2'+(v_2-v_2')-v_2v_2'\cos\psi_2'] \\
\sin\psi_2' \\ 0
\end{array}\right).\label{k2}
\EEA
The velocities of the void are   
\BEA
v_i&=&R \f{dr_v}{dt}\bigg |_{t=t_i}=\beta \eta(t_i) 
\label{velocity-void}
\\
v_i'&=& R'\f{dr'_v}{dt'}\bigg |_{t'=t_i'},
\EEA
where $i=1,2$. From the connection conditions
(\ref{con1}) and (\ref{con2}), and equation (\ref{dt'2}), up to 
order $O(\eta^3)$, $v_i'$ can be calculated as
\BEA
v_i'(t_i)&=&\biggl(1-\f{2}{9}\eta_i^2 \kappa_i\biggr)
\biggl\{\biggl ( 
1+\f{2}{3}\beta \delta_{Hi} \eta_i^2-
\f{2}{9}\delta_{Hi}^2 \eta_i^2
\biggr)\biggl(v_i+\f{2}{3}\eta_i\biggr)
\nonumber
\\
&-&\f{2}{3}\eta_i (1+\delta_{Hi})\biggr\}, 
\label{vi'}
\EEA
where $\eta_i\equiv \eta(t_i)$and
$\delta_{Hi}\equiv \delta_H(t_i), \kappa_i\equiv (1+\delta_{Hi})^2
-(1+\delta(t_i))$.
Using the Friedmann equation, 
the parameters at $t=t_1$ can be written in terms of those
at $t=t_2$ as
\BE
1+\delta_1=\f{H_2^2}{H_1^2}
\biggl(\f{R_2'}{R_1'}\biggr)^3 (1+\delta_2),
\EE
and
\BE
\kappa_1=\f{H_2^2}{H_1^2}
\biggl(\f{R_2'}{R_1'}\biggr)^2 \kappa_2.
\EE
From the geometry of the void in the internal 
comoving frame, the relation between the void radius 
$r_{v1}'$ and $r_{v2}'$ is approximately given by
\footnote{Strictly speaking, equations (\ref{geom1}) and 
(\ref{geom2}) are not
exact, because the photon path is not generally straight 
inside the void. However,
the curvature correction to the photon path is negligible
at the order $O(\eta^3)$ (M\'esz\'aros \& Moln\'ar, 1996).   
}
\BE
r_{v1}' \sin{\psi_1'}\approx r_{v2}' \sin{\psi_2'}. 
\label{geom1}
\EE
The relation between the time $t_1$ and $t_2$
can be written as
\BE
\int_{t_1'}^{t_2'} \f{dt'}{R'(t')} \approx r_{v1}' \cos \psi_1'+r_{v2}' \cos
\psi_2'. \label{geom2}
\EE
The energy loss suffered between times $t_1$ and $t_2$ by a
CMB photon that does not traverse the void is
\BE
\biggl(\f{E_2}{E_1}\biggr)_{\textrm{no void}}=\f{1+z_1}{1+z_2},
\EE
where $z_1$ and $z_2$ are redshift parameters corresponding
to $t_1$ and $t_2$, respectively.
The ratio of the temperature change for photons that traverse the void 
to that for photons that do not traverse the void is
\BE
\f{\Delta T}{T}=\biggl( \f{E_2}{E_1}\biggr )_{\textrm{void}}
\f{1+z_2}{1+z_1}-1. \label{dToverT}
\EE
Equations (\ref{scale}-
\ref{con2}), 
(9),(11),(18)-(30), and (32) can be solved recursively using $\eta_2$ as 
a small parameter.

\begin{figure*}[t]
\includegraphics[width=17cm]{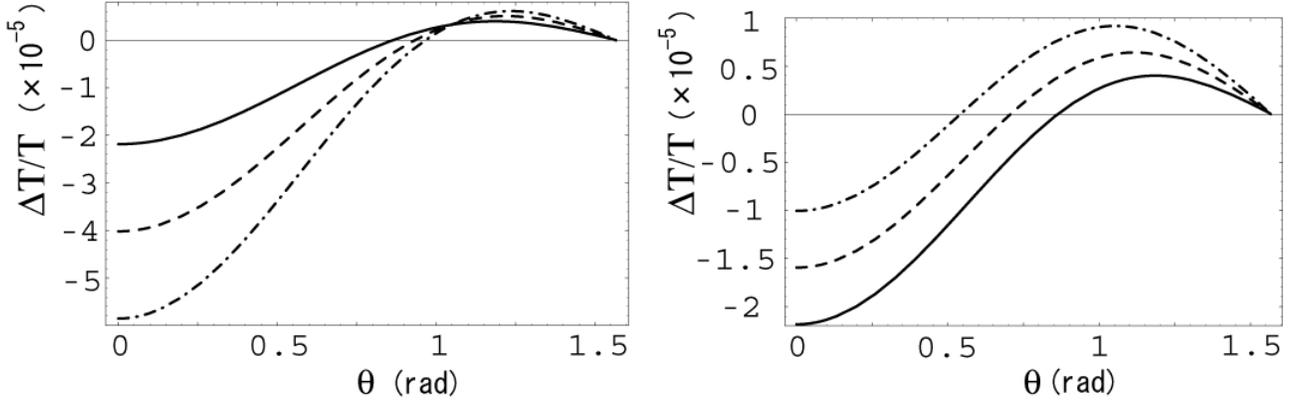}
\caption{The temperature anisotropy for a void 
with $r_v=340h^{-1}$Mpc, $\delta=-0.3$ and the size/distance parameter 
$w=1$ is plotted as a function 
of the subtending angle 
$\theta$ for various parameters that control the second order effects
{\ti{Left}}:($\epsilon,\xi$)=(0,0) (full curve), 
 ($0.1,0$) (dashed curve), and ($0.2,0$) (dashed-dotted curve). 
{\ti{Right}}:($\epsilon,\xi$)=($0,0$) (full curve), 
 ($0,0.05$) (dashed curve), and ($0,0.1$) (dashed-dotted curve).  }
\label{dToverTvoidN.eps}
\end{figure*}
After a lengthy calculation, neglecting the terms 
of order $O(\kappa^2)$, we find
\BEA
\f{\Delta T}{T}&=&\f{8}{81}\eta_2^3 \cos\psi_2 
\biggl( -9\beta(2\delta_H+\delta_H^2-\kappa)-2 \delta_H \kappa
\nonumber
\\ 
&+&
2(\delta_H(5-2 \kappa)+7\delta_H^2+3 \delta_H^3-\kappa)
\cos^2 \psi_2
\biggr), \label{result}
\EEA
where $\delta_H \equiv \delta_H(t_2)$ and 
$\kappa \equiv (1+\delta_H(t_2))^2-(1+\delta(t_2))$.

For $\delta_H=-1$ and $\kappa=0$, we recover the
formula for an empty Minkowski void (Thompson \& 
Vishniac 1987). Note that the formula can also be 
obtained by considering the
hyperbolic vacuum limit: $\delta_H=0$ and $\kappa=1$.
For a quasi-non-linear void, the linear approximation
$\delta_H =-\delta/3 $ holds if the background 
universe is matter-dominated (see Appendix). Then 
equation (\ref{result}) yields
\BE
\f{\Delta T}{T}=- \f{8 }{2187}\delta  \eta_2^3 \cos{\psi_2}
\Bigl(243 \beta -(\delta-42)\delta + \delta (12+\delta) 
\cos(2\psi_2)  \Bigr),
\label{result2}
\EE
where $\delta \equiv \delta(t_2)$. Plugging 
equation (\ref{beta}) for $\Omega=1$ into equation (\ref{result2}), 
we have a simple formula for a compensating void,
\BE
\f{\Delta T}{T}\approx -\f{4 }{2187}\delta^2  \eta_2^3
\cos{\psi_2}(3+24 \cos{2 \psi_2}).
\label{result3}
\EE
Thus, we would observe the dust-filled thin shell void 
as a cold spot surrounded by a slightly hot ring 
in the sky with an anisotropy $\Delta T/T \approx -4
\delta^2 \eta_2^3/81$ toward the center of the void.
To generate an anisotropy $\Delta T/T=10^{-5}$, 
for a given density contrast $\delta_0$ at the present, the 
void radius should be
\BE
r_{v}(z=0)\approx 
\Bigr(\f{\delta_0}{0.2}\Bigr)^{-2/3} 3\times 10^2~~h^{-1}
\textrm {Mpc},
\EE
where $H^{-1}_0=3000 h^{-1}$Mpc. 

From equation (\ref{result3}),
one can see that the terms that are linear in 
$\delta$ cancel each other if the condition $\delta_H=-\delta/3$ 
is imposed. Therefore, we conclude that the 
anisotropy owing to the void is a second-order 
effect. To model the non-linear effects, we introduce two new 
parameters $\epsilon$ and $\xi$ that control the second 
order effect. They are defined as $\delta_H=
-\delta/3-\epsilon \delta^2$ and $\beta=-\delta/6+\xi\delta^2$ 
(Morita et al. 1998, Noh\& Hwang 2005).
One can see in the left panel 
in figure \ref{dToverTvoidN.eps} that 
increasing the value of $\epsilon$ leads to an enhancement in the 
redshift of photons that enter the neighborhood of the center of
the void because the increase in the (absolute) curvature $\kappa$ 
(or equivalently the decrease in the density) allows 
the photons to travel through the void in a shorter time.   
On the other hand, increasing the value of $\xi$ 
leads to an enhancement in the blueshift of photons  
whose trajectories pass near the edge of the void.  This is 
because an increase in the 
expansion rate of the
void shell causes a larger difference between the 
gravitational potentials at the time that the photon enters 
and at the time that the photon leaves. 

To calculate the angular power spectrum, we need the 
angular size $\theta$ of the void. For a void at 
redshift $z$ centered at an angular diameter distance 
$d$ from the Earth, we have
\BE
\psi(w,\theta)=\sin^{-1}(\sin\theta/w) , \label{psi}
\EE
where the size/distance parameter is defined as 
$w \equiv R(z) r_v/d$, and $\theta$ is the angular size of the 
void. From equations (\ref{result}), (\ref{psi}) and non-linear 
parameters $(\epsilon, \xi)$, we
can analytically calculate the angular power of the anisotropy.

\section{Origin of large-angle anomalies}
In this section, we construct a model of 
local voids that can explain the 
observed CMB large-angle anomalies.  We assume that 
the voids are ``local'' at redshifts $z\lesssim 1$.

\subsection{The quadrupole/octopole alignment and planarity}
Tegmark et al. (2003) have found two new anomalies
on very large angular scales 
in the CMB map from the WMAP satellite (see also Schwarz et al. 2004; 
de Oliveira-Costa et al. 2004).
Firstly, the preferred axis of the quadrupole
is aligned to that of the octopole.  
Although, the definition of the ``preferred axis''
of these low multipoles is not unique, a simple
formulation is to think of the CMB temperature fluctuation
as the wave function 
\BE
\f{\Delta T}{T}(\hat{\textbf{n}})\equiv \Psi(\hat{\textbf{n}})
\EE
and find the axis $\hat{\textbf{n}}$  
around which the angular momentum dispersion 
\BE
s(\hat{\textbf{n}})\equiv \langle \Psi|(\hat{\textbf{n}}\cdot \textrm{L})^2  | 
\Psi\rangle=\sum_{m} m^2 |a_{lm}(\hat{\textbf{n}})|^2 
\EE
is maximized (de Oliveira-Costa et al. 2004). 
The interval between the unit vectors $\hat{\textbf{n}}_1$ 
and $\hat{\textbf{n}}_2$ corresponding to the 
``preferred axes'' for the quadrupole and octopole is within 
$\sim 10^\circ$(de Oliveira-Costa et al. 2004). 

Another anomaly is the unusual 
planarity of the octopole. To measure the 
octopole planarity, we define the octopole 
concentration parameter ${\cal{T}}$ as the percentages of 
the octopole power that can be attributed to the modes
with $|m|=3$,
\BE
{\cal{T}}(\hat{\textbf{n}})\equiv \f{|a_{3 -3}(\hat{\textbf{n}})|^2+
|a_{3 3}(\hat{\textbf{n}})|^2}{\Sigma_{m=-3}^3 
|a_{3 m}(\hat{\textbf{n}})|^2}.
\EE 
For the TOE map, choosing the preferred direction in which the
angular dispersion around $z$-axis is maximal, the octopole
concentration parameter is
${\cal{T}}=94\%$. The relation of the 
octopole planarity and the concentration
to the $|m|=3$ modes is obvious because the wavenumber for the
fluctuation in the 
spherical harmonic $Y_{3,-3}$ or $Y_{3,3}$ as a function of the
polar angle $\theta $ is $k\sim 1$, corresponding to a half-wave
on the meridian.

To explain the quadrupole/octopole alignment and the octopole 
planarity, we consider {\ti{a
pair of identical local voids with a 
separation angle between the center 
$\theta_s=50^\circ-60^\circ$ at 
$\hat{\textbf{n}}_1$ and $\hat{\textbf{n}}_2$}}. 
If the contribution of these cold spots
is dominant over the CMB in the quadrupole and the octopole, 
then the preferred axis for both multipoles
is parallel to $\hat{\textbf{n}}_1-\hat{\textbf{n}}_2$
for which the angular momentum dispersion parallel
to the rotation axis is maximum.   
Therefore, the quadrupole/octopole alignment is 
naturally explained in such a configuration. 
In the coordinate system in which the 
$z-$axis is aligned to the preferred axis, the 
equator lies on the equidistant point between
the two identical voids. Then the wavenumber of 
the fluctuation owing to the two cold spots 
is roughly $k\sim 1$ for the separation angle 
$\theta_s\sim 60^\circ$ between 
the void. As shown in figure \ref{separation.eps},  
we find a ${\cal{T}}=99.9 \%$ concentration for 
a separation angle $\theta_s=53^\circ$ in the 
preferred direction
independent of the size/distance ratio $w=R r_v/d$. 

\BF
\includegraphics[width=7cm]{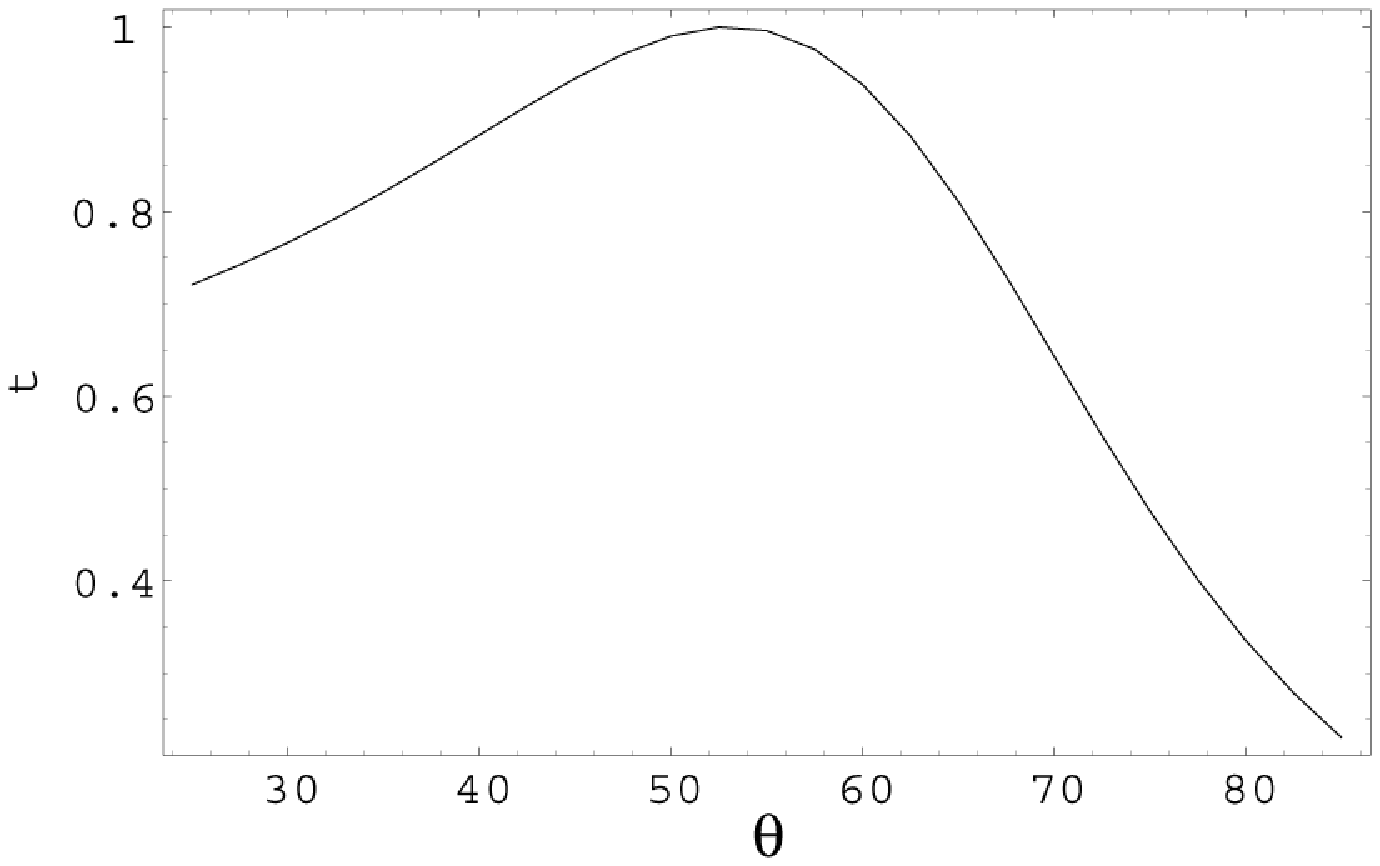}[t]
\caption{The octopole 
concentration parameter $t$ as a function of the separation
angle $\theta$ (in degree) between the pair of identical voids.}
\label{separation.eps}
\EF

\begin{figure*}

\epsscale{1.2}
\plotone{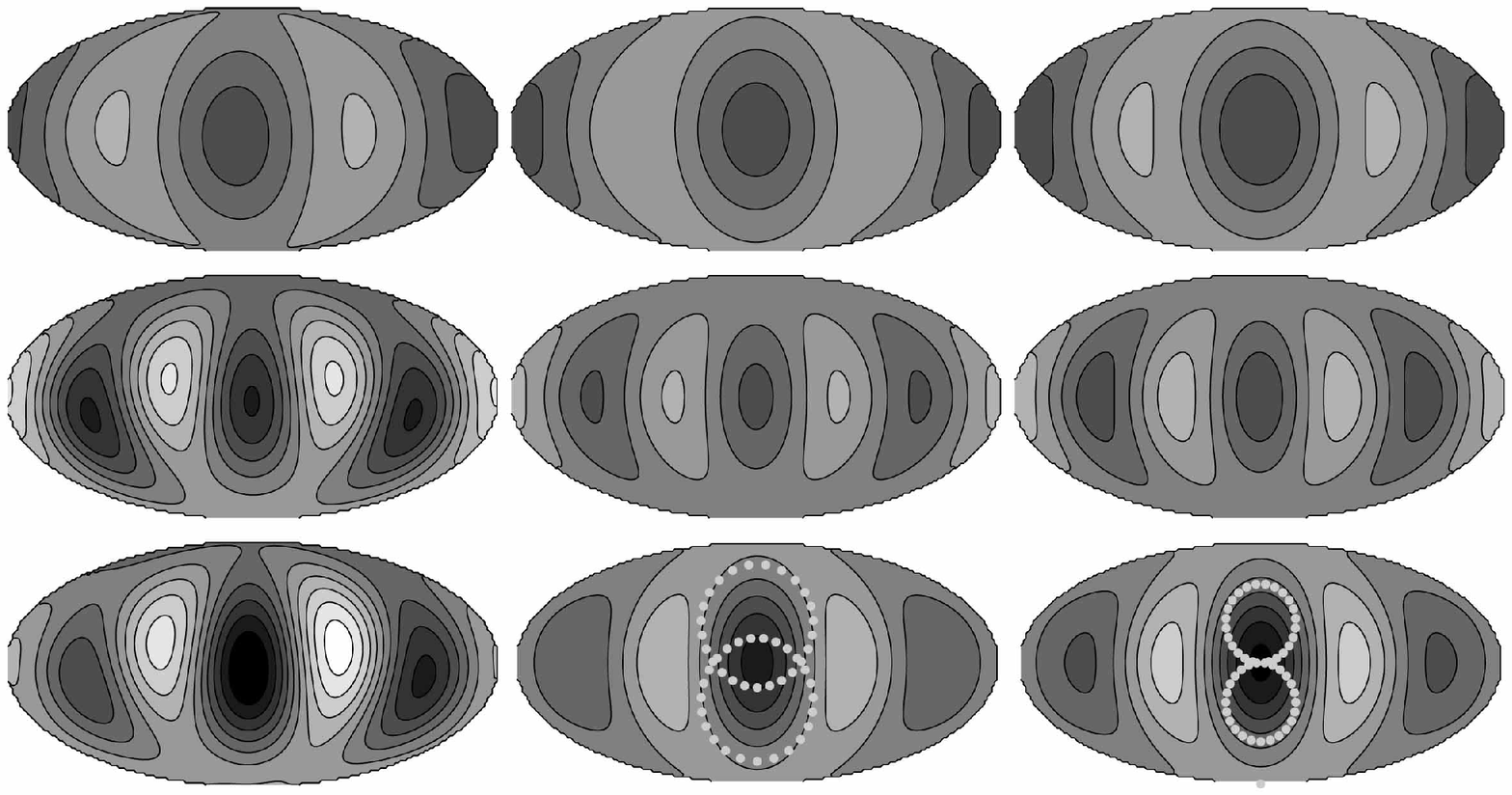}
\caption{The Mollweide projection maps 
for the cleaned CMB (Tegmark et al. 2003)
 (left) and those for a
pair of identical voids with the size/distance ratio 
$w=0.9$ and the non-linear parameters $(\epsilon,\xi)=(0,0)$ (middle) 
and $w=0.47$, $(\epsilon,\xi)=(0.3,0)$ (right). 
The separation angles of the voids and the void radius are 
assumed to be $\theta_s=50^{o}$, $r_v=340 h^{-1}$Mpc, respectively. 
The gray scale denotes the temperature fluctuations 
in which the maximum absolute value is set to $50\mu$K. 
The north pole is aligned to the $z-$axis for which the
angular dispersion $s(\hat{\textbf n})$ of the quadrupole
plus the octopole around the $z-$axis is maximal in the direction
$(l,b)=(-110^\circ,60^\circ)$. The coordinate center is located 
at $(l,b)=(-30^\circ,-30^\circ)$.  
{\it{Top}}: the quadrupole maps. {\it{Middle}}:the octopole maps.
{\it{Bottom}}:the quadrupole plus the octopole maps. The boundaries 
of the void hemispheres are denoted by small 
light-gray disks.    } \label{skymap.eps}
\end{figure*}

\begin{figure*}
\epsscale{0.7}
\plotone{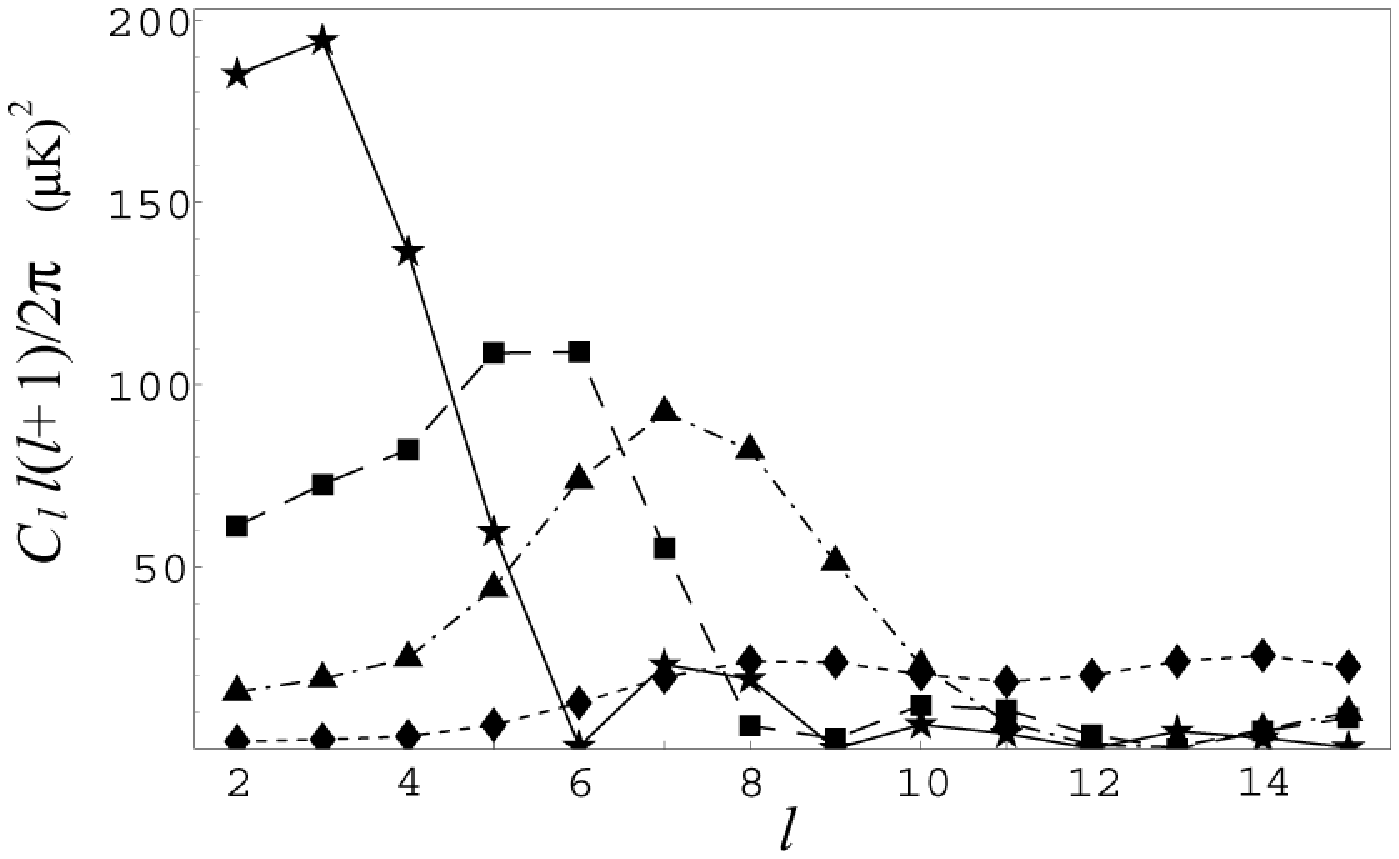}
\caption{Plots of the 
angular power $T_l=C_l l(l+1)/2 \pi$ owing to a pair of
 identical voids with $(\epsilon,\xi)=(0,0)$
as a function of 
$l$ for different locations 
$w=0.9$ (star), $w=0.7$ (box), $w=0.5$ (triangle), 
$w=0.3$ (diamond) for the separation angle $\theta=50^\circ$. We assumed
$\delta=-0.3$. and $r_v=340 h^{-1}$Mpc. The distance to the 
void is given by $d\sim r_v/w$ at $z \sim 0$. }
\label{Tlvoid.eps}
\end{figure*}
We can see in figure \ref{skymap.eps} that the correlation between 
the CMB quadrupole and the CMB octopole is large 
only for a cold spot centered 
at $(l_c,b_c)=(-30^\circ,-30^\circ)$ and a  
hot spot centered at $(l_h,b_h)=(-100^\circ,-20^\circ)$,
suggesting that the alignment is due to these local
fluctuations. Although the hot spot might be
related to residual Galactic emission, it would
be difficult to explain the cold spot in this way. 
For comparison,
we have simulated the temperature anisotropies for
a pair of identical voids with radius $r_v=340 h^{-1}$Mpc for 
parameters $w=0.9$, $(\epsilon,\xi)=(0,0)$ and 
$w=0.47$,$(\epsilon,\xi)=(0.3,0)$.
As shown in figure \ref{skymap.eps}, 
we can see that the void model can generate 
a similar cold spot as in the observed CMB. The preferred axis is 
parallel to the vector that connects the centers of the voids.

\subsection{The unusually cold spot}

Using a wavelet technique, Vielva et al. (2004) detected 
an unusually cold spot in the Galactic southern 
hemisphere at $(l,b)=(-153^\circ, -59^\circ)$. 
The significance in the form of kurtosis on $\sim 4^\circ$ scales
is at the $3\sigma $ (a fluke at the one in $\sim 500$) level 
relative to the simulation of Gaussian fluctuations.  
Such a cold spot can be naturally explained by a presence of 
a large void at $z\sim 1$ in the direction of the line of sight
to the cold spot. Assuming the 
Einstein-de-Sitter background, for 
a density contrast $\delta\sim -0.3$ at $z \sim 1$ 
and comoving radius $r_v \sim 210~h^{-1}$ Mpc, the temperature
anisotropy would be $\Delta T/T= -4\delta^2 \eta^3/81\sim -10^{-5}$ 
in the direction of the void. The corresponding angular 
diameter distance at $z=1.0$ is $d=0.29~H_0^{-1}$.
Then the angular radius of the void would be
$\theta_v\sim 7^\circ$ and $w\sim 0.17$. 
For the flat-$\Lambda$ universe with $\Omega_0=0.3$, 
the corresponding angular 
diameter distance at $z=1.0$ is $d=0.39~H_0^{-1}$, then
the angular radius of the void would be 
$\theta_v \sim 5^\circ$. 
\subsection{The north-south asymmetry}
Another anomaly is the  
so-called north-south asymmetry: the
systematic difference in the amplitudes of fluctuations 
on opposing hemispheres. Eriksen et al. (2004)
 first pointed out the asymmetry in the power 
for multipoles $5<l<40$ between the two hemispheres
in which the ``north pole'' is located at 
$(l, b)\sim(60^\circ,10^\circ)$.
Subsequent analyses (Hansen et al. 2004)  
confirmed their result. 

In our void model, the asymmetry is naturally explained
by the asymmetric distribution of voids on the sky.
As shown in figure \ref{Tlvoid.eps}, the pair of voids 
at $w=0.9$ contributes for multipoles $l=2-5$ and 
voids at $w<0.7$ contributes for higher multipoles $l>5$
because the corresponding angular scale is $l\sim \pi/\tan^{-1}w $.
The number $N$ of voids that contribute to the CMB large-angle power
should be small because the asymmetry in the power 
falls off as $1/\sqrt{N}$. 

\subsection{The low quadrupole}
The origin of the anomalously low quadrupole has been 
extensively discussed since it was first observed by
COBE/DMR (Smoot et al. 1992). The WMAP team argued that 
the low quadrupole requires a selection  at the one 
in 143 level (Bennett et al. 2003). However, recent
studies based on  Bayesian or  frequentist analyses 
have shown that the statistical significance
is actually much less than  previously claimed 
(Tegmark et al. 2003, Efstathiou 2004). 
The key issue is the modeling of the 
foreground properties.  Even though the observed power is low, 
most of the power may be hidden inside the Galactic 
foregrounds. The uncertainty in the amplitude of the 
cosmological background near the Galactic
plane will generally lower the statistical significance of the 
observed power.

In a similar manner, the statistical significance 
of the low quadrupole can be further reduced 
in our void model. Consider a pair of voids
with identical angular radius $\theta_v=\sin^{-1} w $. 
Then the portion of the 
sky area masked by the pair of voids is approximately 
$f\sim 1-\cos \theta_v$.
For $w=0.7$, we have $\theta_v=44^\circ$ and $f\sim 25\%$ 
which is larger than 
the portion of the WMAP sky masks such as Kp0, Kp2 and Kp4
 (Bennett et al. 2003).   
Therefore, we expect a further reduction in the statistical 
significance of the quadrupole if the pair of the 
voids is located sufficiently near to the Earth.

However, one may object that the amplitude enhancement would require 
an unusual cancellation between the intrinsic and the 
void-induced quadrupoles. This difficulty may be circumvented
by the following argument. In the standard $\Lambda$CDM model, 
the adiabatic large-angle anisotropy $l<5$ owing to the
``late'' integrated Sachs-Wolfe (ISW) effect (redshift effect 
for a void) partially
cancels the ``ordinary'' Sachs-Wolfe (OSW) 
effect (Hu \& Sugiyama 1995).
As we have seen in the previous section, the photon
that traverses a point near  the center is  redshifted.
Thus, the presence of the void may provide a
further cancellation between the ISW effect and the OSW effect.
A  quantitative estimate is deferred to a later analysis.

\section{Summary and Discussion}
In this paper, we have explored the temperature 
anisotropy owing to dust-filled homogeneous local
voids at $z \lesssim 1$ in the flat FRW universe. 
Local voids with a comoving radius 
$r_v\sim3 \times 10^2 h^{-1}$ Mpc and a 
density contrast $\delta=-0.3$ will give a 
cold spot with anisotropy $\Delta T/T \sim -10^{-5}$
surrounded by a slightly hotter ring. 
A pair of such compensated voids at 
distance $\sim 4\times 10^2 h^{-1}$Mpc
whose equidistant point in the 
direction $(l,b)=(-30^\circ,-30^\circ)$ can
explain the quadrupole/octopole alignment and the octopole planarity,
and the void at $z\sim 1$ in the 
direction $(l,b)=(-153^\circ, -59^\circ)$
can be the origin of the unusually cold spot. 
We also find that the north-south asymmetry can be explained by the
asymmetric distribution of these local voids. 

As we have shown, the temperature anisotropy owing to a dust-filled 
void in the Einstein-de Sitter background is 
proportional to $\delta^2 r_v^3$. Assuming the CDM power spectrum
for the linear perturbation, the expected temperature 
anisotropy owing to the void is maximum at the scale $r_v 
\sim 100h^{-1}$Mpc, because $\delta \propto r_v^{-2}$ for the linear
regime whereas $\delta \propto r_v^{-1/3}$ for the non-linear
regime. Thus the contribution from the single
non-empty voids at the scale $r_v \sim 100h^{-1}$Mpc is more important than the empty voids at smaller scales. 

For simplicity, we have not considered the effect of the  
cosmological constant. 
However, we expect that the change in 
the deceleration parameter will not dramatically alter the order
of magnitude 
of the anisotropy although the presence of the linear ISW effect
$\sim O(\delta r_v^3)$ may somehow affect the factor. The amplitude
owing to the second-order effect would
be $\sim O(\delta^2 r_v^3)$. Even if 
the density contrast $\delta$ is small, 
the correlation between the linear ISW term 
and the second order term can enhance the observed 
signal (Tomita 2005a,2005b). The extension of our analysis to 
a flat-$\Lambda$ FRW background will be presented
in the forthcoming paper (K.T. Inoue \& J. Silk 2006, in preparation).

We did not go into details about the 
non-linear evolution of the voids. Our 
thin-shell approximation should provide a
good description of voids that are sufficiently smaller than the 
Hubble radius in the late stages of evolution. 
However, in the middle stages, the density fluctuation and
the peculiar velocity of the shell may depend
on the initial configuration
of the density and the velocity perturbation. 
As we have seen, the temperature anisotropy owing to 
a quasi-non-linear void is a 
second-order effect (see also Seljak, 1996; Cooray 2002;
Tomita 2005a, 2005b), 
which may significantly enhance
the anisotropy. Therefore, we need a more elaborate calculation 
 to make a more precise prediction about the anisotropy
attributed to the local voids. 

We have not addressed the issue of void origin. One may find
discussions of this in previous treatments of the impact of voids on
small angular scale CMB observations
(Griffiths et al. 2003).  The voids postulated here
represent the large-radius tail of the void distribution discussed in
these earlier studies. They represent a non-Gaussian fluctuation
distribution that is linearly superimposed on the usual
scale-invariant spectrum of Gaussian adiabatic density
fluctuations. Mathis et al. (2004) discussed the implications of a
distribution of primordial voids for galaxy and cluster
formation. There are similar implications for the large voids that we
have postulated if these are part of a statistical ensemble of
primordial voids. In rare regions, associated with the thin shell
surrounding the voids, cluster and galaxy formation will be enhanced.
In principle this will be detectable via massive clusters and galaxies
at redshifts significantly larger than expected in the purely Gaussian
model. Examples  include detection of massive galaxies and galaxy
clusters at high $z$ and massive clusters with high concentrations of
dark matter. All of these cases, if confirmed, would 
 represent potential difficulties for the
standard model of primordial Gaussian fluctuations.

Another implication unique to the very large $\sim 300~h^{-1}$ Mpc
voids postulated here concerns increased dispersion in the Hubble
constant as measured both in different directions and at different
redshifts (Tomita 2001). The dispersion 
will be positively skewed since voids tend to 
generate a larger Hubble constant. Moreover, the global value of the
Hubble constant will be lowered.  The effect could be as large as
$\sim 10\%.$

So far, the large scale voids we have assumed have not been detected.
In the future, the ongoing projects such as the 
6dF galaxy survey (Jones et al. 2004) may 
detect our postulated voids (underdensity region).

\acknowledgments

We thank K. Tomita, N. Sakai, G. Starkman, S. Zaroubi and N. Sugiyama
for useful discussions and comments. This work is in 
part supported by a Grant-in-Aid for
Young Scientists (17740159) of the Ministry of Education, Culture,
Sports, Science and Technology in Japan.

\appendix
\section{Perturbation of Hubble parameter}
Let us write the Friedmann equation 
at a point $\x$ in a perturbed universe in terms of
the scale factor $a$, the Hubble parameter $H$, the 
gravitational constant $G$, and the curvature $K$ as 
\BE
H^2(\x,t)=\f{8 \pi G}{3}\rho(\x,t)-\f{K(\x,t)}{a^2(t)}
\label{Friedmann}
\EE
where the Hubble parameter $H$ is defined on a comoving slice.
>From the perturbed equation (\ref{Friedmann}) in 
the Fourier space, for the flat FRW background without $\Lambda$, 
we have
\BE
2 H \delta H_\K=\f{8 \pi G}{3}\delta 
\rho_\K-\f{2}{3}\biggl(\f{k}{a}  \biggr)^2 {\cal{R}}_\K, 
\label{Friedmann2}
\EE
where the curvature perturbation ${\cal{R}}$ is defined as 
\BE
\delta K_\K=\f{2}{3} k^2 {\cal{R}}_\K.
\EE 
From the Poisson equation, 
the Newtonian gravitational potential
$\Phi_\K$ is written in terms of the density contrast $\delta_\K$ 
as
\BE
\delta_\K=-\f{2}{3}\biggl(\f{k}{a H}\biggr)^2 \Phi_\K. 
\label{deltaK}
\EE
For a constant equation of state $w\neq-1$, the continuity  
equation gives a relation between 
the gravitational potential $\Phi_\K$ and 
the curvature perturbation $\cal{R}_\K$ as (Liddle \&
Lyth 2000)
\BE
\Phi_\K=-\f{3+3 w}{5+ 3 w} \cal{R}_\K.
\label{PhiK}
\EE
From equations (\ref{Friedmann2}), (\ref{deltaK}), and (\ref{PhiK}),
we have
\BE
\f{\delta H_\K}{H}=-\f{\delta_\K}{3(1+w)}.
\EE
Thus, for the matter-dominated flat FRW universe $w=0$, the 
Hubble parameter contrast is written as 
$\delta_H=-\delta/3$ in the linear order.

\end{document}